\documentclass{ws-procs975x65}

\def\beq{\begin{equation}}
\def\eeq{\end{equation}}

\begin{document}

\title{Experimental Tests of Local Cosmological Expansion Rates}
\author{Allan Widom$^*$ and John Swain}

\address{Physics Department, Northeastern University \\
Boston, MA, 02115, USA\\
$^*$E-mail: allan\_widom@yahoo.com\\
john.swain@cern.ch}

\author{Yogendra Srivastava}

\address{Physics Department, University of Perugia \\
Perugia, Italy\\
E-mail: yogendra.srivastava@gmail.com}

\begin{abstract}
Cosmological expansion on a local scale is usually neglected in part
due to its smallness, and in part due to components of bound systems
(especially those bound by non-gravitational forces such as atoms
and nuclei) not following geodesics in the cosmological metric. However, it is interesting to 
ask whether or not experimental tests of cosmological expansion on
a local scale (well within our own galaxy) might be experimentally
accessible in some way. We point out, using the Pioneer satellites as
an example,  that current satellite technology allows
for this possibility within time scales less than one human lifetime.
\end{abstract}

\keywords{cosmology, local Hubble expansion, Pioneer}

\bodymatter


\section{Introduction \label{intro}}

The electromagnetic waves sent into our solar system by the Pioneer 
satellites 10 and 11 exhibit frequency shifts which vary with
time\cite{Anderson:1001}. Such frequency shifts were parameterized as 
accelerations dominated by the gravitational attraction between the solar system and 
the satellites. An anomalous acceleration remnant was reported which cannot 
be explained by the usual Newtonian gravitational field produced by the sun 
at the positions of the satellites. 

To see what was involved, consider the velocity 
\begin{math} {\bf v}(t) \end{math} of a satellite moving away from the sun 
slowly into the milky way galaxy at a non-relativistic speed 
\begin{math} v(t)=|{\bf v}(t)|\ll c \end{math}. What was expected was a 
Doppler shifted frequency of the electromagnetic wave which could be 
employed as an accelerometer;   
\begin{equation}
\omega (t)=\omega_0\left(1-\frac{v(t)}{c}+\cdots \right) 
\ \ \ \ \ \Rightarrow 
\ \ \ \ \ \frac{\dot{\omega}(t)}{\omega(t)} 
=-\frac{\dot{v}(t)}{c}+\cdots \ \ .  
\label{intro1}
\end{equation}
The remnant experimental acceleration anomaly 
\begin{math} -{\cal A}_{\infty }  \end{math} was 
reported\cite{Anderson:1001} as   
\begin{equation}
{\cal A}_{\infty}=\lim_{t\to \infty } \dot{v}(t) 
\approx 7.5\times 10^{-8} \left[\rm cm \over sec^2\right], 
\label{intro2}
\end{equation}
wherein \begin{math} \infty  \end{math} denotes a time sufficiently large 
so that the gravitational acceleration towards the sun should have been 
virtually zero at least from the viewpoint of Newtonian gravitational theory. 
The experimental result reported in Eq.(\ref{intro2}) was thereby thought by 
many to be an anomaly, if not an experimental error. For example, 
Turyshev {\em et al.} suggest a thermal origin\cite{Turyshev} of the
data, but also acknowledge that open questions remain.

Regardless of whether or not one considers the Pioneer anomaly to be
resolved (that is, explained by experimental factors now accounted for) or not, the
important point we raise is that the level of precision is such that one
can think about experimental tests of cosmological expansion on distance
scales which can be reached by earth-launched satellites within one
human lifetime.

\section{Uniform Cosmological Metrics \label{um}}

The standard uniform cosmological metric may be written via the proper time 
\begin{equation} 
c^2 d\tau^2 = c^2 dt^2 -a(t)^2 d\ell ^2\ \ \ \Rightarrow 
\ \ \ d\ell ^2=
d\chi^2 +f_\kappa ^2 (\chi )\left[d\theta^2 +\sin^2\theta d\phi^2 \right], 
\label{um1}
\end{equation}
wherein \begin{math} \kappa =0,\pm 1,
\ f_{+1} (\chi )=\sin \chi , \ f_0(\chi )=\chi 
\ {\rm and}\  f_{-1} (\chi )=\sinh \chi  \end{math}.
The Hubble expansion rate is thereby 
\begin{equation} 
{\cal H}=\frac{\dot{a}}{a}\ .  
\label{um2}
\end{equation}
A central issue in cosmology is to describe how the uniform metric in 
Eq.(\ref{um1}) can be employed to describe a very non-uniform 
distribution\cite{Springel:2006} of observable gravitational masses.
We follow convention in what follows, making no special claims
about the possible non-uniformity or indeed presence of dark
matter and dark energy, and simply take the standard cosmological metrics
(which are applied with or without dark matter or energy) to be
good approximations to parametrize cosmological metrics.

\section{Local Hubble Expansion \label{lhe}}

According to general relativity and the standard cosmological 
models\cite{Landau:1999,Weinberg:2008}, the universe is presently 
expanding at a Hubble rate 
\begin{math} {\cal H} \end{math} measured in part by the rate at which 
frequencies of electromagnetic radiation change with time 
\begin{equation}
\frac{\dot{\omega }}{\omega }=-{\cal H}=-\frac{1}{\cal T}\ ,
\label{lhe1}
\end{equation}
where ${\cal T}$  is the Hubble time.

If a local Hubble expansion does exist, then the data should be 
parametrized by the assignment    
\begin{equation}
\frac{{\cal A}_\infty}{c}={\cal H}=\frac{1}{\cal T} 
\ \ \ \ \Rightarrow 
\ \ \ \ {\cal T}=\frac{c}{{\cal A}_\infty} 
\label{lhe2}
\end{equation}

If one uses the Pioneer value (assuming it to be
a real effect rather than an experimental artefact) of ${\cal A}_\infty \approx 7.5\times 10^{-8} \left[\rm cm \over sec^2\right]$,
together with the presence of Hubble expansion rate on
a local length scale of the order of the Pioneer path length
one obtains

\begin{equation}
{\cal T}=\frac{c}{{\cal A}_\infty} \approx 12.7\times 10^9\ {\rm year}. 
\label{lhe3}
\end{equation}

Rather remarkably, this is a numerical value for the Hubble time in good agreement 
with other measurements of this time which have been 
reviewed\cite{Weinberg:2008}. 

Thus, if one assumes the Pioneer acceleration to be a real effect, the gravitational acceleration 
toward the sun appears indeed virtually zero where and when it should be 
according to Newtonian gravity and the general relativistic
Hubble expansion would appear to explain it.

Conversely, if the original apparent observed Pioneer acceleration is explained
by experimental errors and actually zero, then this provides evidence
against a local cosmological expansion.

Regardless of the interpretation of the data -- and this is the key point of this paper --
the precision which can be obtained from data from missions like the
Pioneer ones appears sufficient to allow questions about local cosmological
expansion to be experimentally tested, without having to wait for cosmological
time scales, and indeed, well within the lifespan of a human being.

\section{Conclusion \label{conc}}

As noted in the abstract, cosmological expansion on a local scale is usually neglected in part
due to its smallness, and in part due to components of bound systems
not following geodesics in the cosmological metric. Nevertheless, as physics
is an experimental subject, it is interesting to ask whether or not tests can be
made within our cosmic environment using probes which we understand
and can control.

We show, using the Pioneer satellites as
an example,  that current satellite technology allows
for this possibility within time scales less than one human lifetime. Indeed the
data, depending on whether or not the ``anomaly'' is considered to have been resolved,
may already be good enough to start to address such questions.

More data and other experiments would
be welcome (and indeed may be accessible through further analysis of Pioneer data
and via the Deep Space Network \cite{Turyshev}). At the very least, the results here indicate that tests of local
Hubble expansion at relatively small cosmological distances are experimentally
feasible. 

\section*{Acknowledgments}

J. S. would like to thank the United States National Science Foundation for support under PHY-1205845.

\end{document}